\documentclass[12pt,draftcls,onecolumn]{IEEEtran}
\usepackage{graphicx}
\usepackage{amsmath}
\usepackage{amsfonts}
\usepackage[nomarkers,nofiglist,notablist]{endfloat}
\let\MYoriglatexcaption\caption
\renewcommand{\caption}[2][\relax]{\MYoriglatexcaption[#2]{#2}}

\newtheorem{theorem}{Theorem}
\newtheorem{lemma}{Lemma}

\newtheorem{proposition}{Proposition}
\newtheorem{corollary}{Corollary}

\renewcommand{\eqref}[1]{(\ref{#1})}

\begin{document}

\title{Protocol Coding through Reordering of User Resources, Part I:
Capacity Results}
\author{\begin{tabular}{c}
Petar~Popovski$^*$ and Zoran~Utkovski$^{\dagger}$\\
$*$ Department of Electronic Systems, Aalborg University, Denmark \\
$\dagger$ Institute of Information Technology, University of Ulm, Germany \\
Email: petarp@es.aau.dk,  zoran.utkovski@uni-ulm.de\\
\end{tabular}}

\maketitle

\begin{abstract}
The vast existing wireless infrastructure features a variety of 
systems and standards. It is of 
significant practical value to introduce new features and devices 
without changing the physical layer/hardware infrastructure, but 
upgrade it only in software. A way to achieve it is to apply \emph{protocol coding}: encode information in the actions taken by a certain (existing) communication protocol. In this work we investigate strategies for 
protocol coding via combinatorial ordering of the labelled user
resources (packets, channels) in an existing, \emph{primary system}. Such a protocol coding introduces a new \emph{secondary communication channel} in the existing system, which has been considered in the prior work exclusively in a steganographic context. Instead, we focus on the use of secondary channel 
for reliable communication with newly introduced secondary devices, that are low-complexity versions of the primary devices, capable only to decode the robustly encoded header information in the primary signals. We introduce a suitable communication model, capable to capture the constraints that the primary system operation puts on protocol coding. We have derived the capacity of the secondary channel under arbitrary error models. The insights from the information--theoretic analysis are used in Part II of this work to design practical error--correcting mechanisms for secondary channels with protocol coding.

\end{abstract}

\section{Introduction}
\subsection{Motivation and Initial Observations}

After two decades of explosive growth, the starting point for
wireless innovation is changed. With the vast amount of deployed infrastructure and variety of existing systems, it is of 
significant practical value to introduce new features
without changing the physical layer/hardware of the infrastructure, but 
only upgrade it in software. This can be achieved by a suitable, backward--compatible upgrade of the communication protocols. We use the term
\emph{protocol coding} to refer to techniques that
convey information by modulating the actions of a communication protocol.

\begin{figure}
  \centering
  \includegraphics[width=7cm]{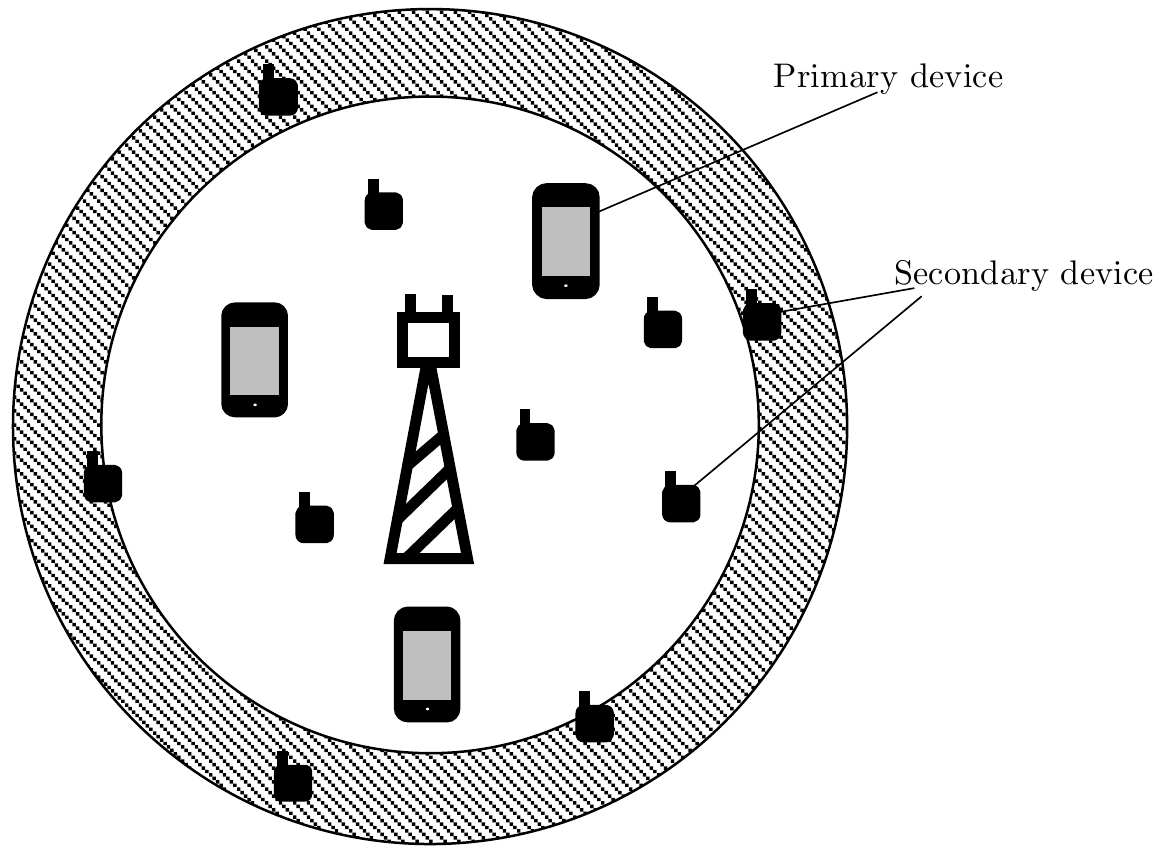}\\
  \caption{Illustration of a secondary communication through protocol coding in cellular systems. A primary device can decode
  any information sent by the base station,
  while the secondary device has a limited functionality can
  only decode the information sent by protocol coding.
  The range of the primary communication system
  (white circle) is smaller than the range of the
  secondary information (shaded circle).}\label{fig:cellular}
\end{figure}

Consider the example on Fig.~\ref{fig:cellular}, where a cellular base station (BS) a group of \emph{primary terminals} in its range. It is assumed that the cellular system is frame--based (WiMax~\cite{WiMax_book}, LTE~\cite{LTE}, etc.). The metadata contained in the frame
header informs the terminals
how to receive/interpret the actual data that follows. The frame header is commonly encoded more robustly
compared to the data, such that it can be reliably received in an area that is larger than the nominal  coverage
area, as depicted on Fig.~\ref{fig:cellular}. In such a context, while still using the same infrastructure, we can introduce new \emph{secondary devices}, which
are able to operate in the extended coverage area. These can be
e.~g. machine-type devices \cite{ref:M2M}, such as sensors or
actuators, that are controlled by the cellular BS. The secondary
devices are simple and have a limited functionality, capable
to decode only the frame header, but not the complex high--rate codebooks used for data. The main idea is that BS can send
information to the secondary devices in the frame header. However,
one could immediately object that the frame header carries important metadata that cannot be changed arbitrarily. The BS
decides how to schedule the primary users based on certain QoS
criterion. Nevertheless, there could be still freedom to rearrange the headers and thereby send information to the secondary devices. To illustrate this point,
assume that there are two OFDMA channels, 1 and 2, defined in a
diversity mode~\cite{WiMax_book}, such that if a user Alice is
scheduled in a given frame, it is irrelevant whether it is
assigned to channel 1 or 2. Hence, if BS schedules Alice
and Bob in a given frame, then it can encode $1-$bit secondary information
as follows: allocating Alice to channel
1 and Bob to channel 2 is a bit value 0, otherwise it is a bit
value 1. Taking this simple example further, let there be three OFDMA channels, but still only two users, Alice and Bob. In a given frame, each of them can get from $0$ up to $3$ channels assigned, which is decided by the primary scheduling criterion; the secondary transmitter can encode information by assigning these channels to Alice/Bob in a particular way. If there are $2 (1)$ packets for Alice (Bob), they can be assigned in $3$ possible ways and in that particular frame, $\log_2(3)$ secondary bits can be sent. However, if all $3$ packets are addressed to Alice, no secondary information can be sent in that frame. This variable amount of information due to the primary operation is the crux of the communication model considered in this work. 

The objective of this and the companion paper~\cite{ref:Part2} is
to investigate the fundamental properties of communication systems
that use protocol coding to send information, under
restrictions imposed by a primary system. The secondary information is encoded in the ordering of labelled resources (packets, channels) of the primary (legacy) users. In this paper
we introduce a suitable communication model that can capture the
restrictions imposed by the primary system. The model captures the key feature of a secondary communication: in a given scheduling epoch, the primary system decides which packets/users to send data to, while secondary information can be sent by only rearranging these packets. Each primary packet is subject to an error (e. g. erasure), which induces a corresponding error model for secondary communication. In this paper we analyze the model
using information-theoretic tools and obtain capacity--achieving
communication strategies, which we then apply in Part II of the work to obtain practical encoding strategies.

\subsection{Related Work and Contributions}

Protocol coding can appear in many flavors. An early work that
mentions the possibility to send data by modulating the random
access protocol is~\cite{Massey}, but in a rather ``negative''
context, since the model used \emph{explicitly prohibits} to
decide the protocol actions based on user data. The seminal
work~\cite{Verdu} uses a form of protocol coding: the information
is modulated in the arrival times of data packets. More recent
works on possible encoding of information in relaying scenarios
through \textit{protocol--level} choice of whether to transmit or
receive is presented in~\cite{Kramer_relay}~\cite{Lutz}
and~\cite{PetarGC10}. At a conceptual level, protocol coding
bridges information theory and networking~\cite{Ephremides}. 
The idea of communication based on
packet reordering is not new per se and has been presented in the context of covert
channels \cite{Ashan} \cite{Chakinala} \cite{AlAtawy}. However, the big difference with our work is that our objective is not steganographic, but rather what kind of 
communication strategies can be used when the degrees of freedom for secondary communication are limited by a certain (random) process in the primary system. The practical coding strategies are related to the
frequency permutation arrays for power line
communications~\cite{HanVinck, PermCodesDesign}.

Preliminary results of this work have appeared in~\cite{Petar} and~\cite{ZoranAllerton}. In \cite{Petar} we have introduced the notion of a secondary channel and sketched of the communication strategies when the primary packets are subject to an erasure channel, while in \cite{ZoranAllerton} we treated the case when the error model for the primary packets is represented by a Z--channel. In this paper we devise capacity--achieving strategies for arbitrary error model incurred on the primary packets and provide the detailed proofs. We first show that our communication model is related to the model of Shannon for channels with causal side information at the transmitter (CSIT)~\cite{ref:ShannonCSIT}. We then develop a new framework for computing the secondary capacity, which leads us to explicit specification of the communication strategies that are applied to convolutional codes in Part II~\cite{ref:Part2}.

\section{System Model}
\label{sec:SystemModel}

\subsection{Communication Scenario}

\begin{figure}
  \centering
  \includegraphics{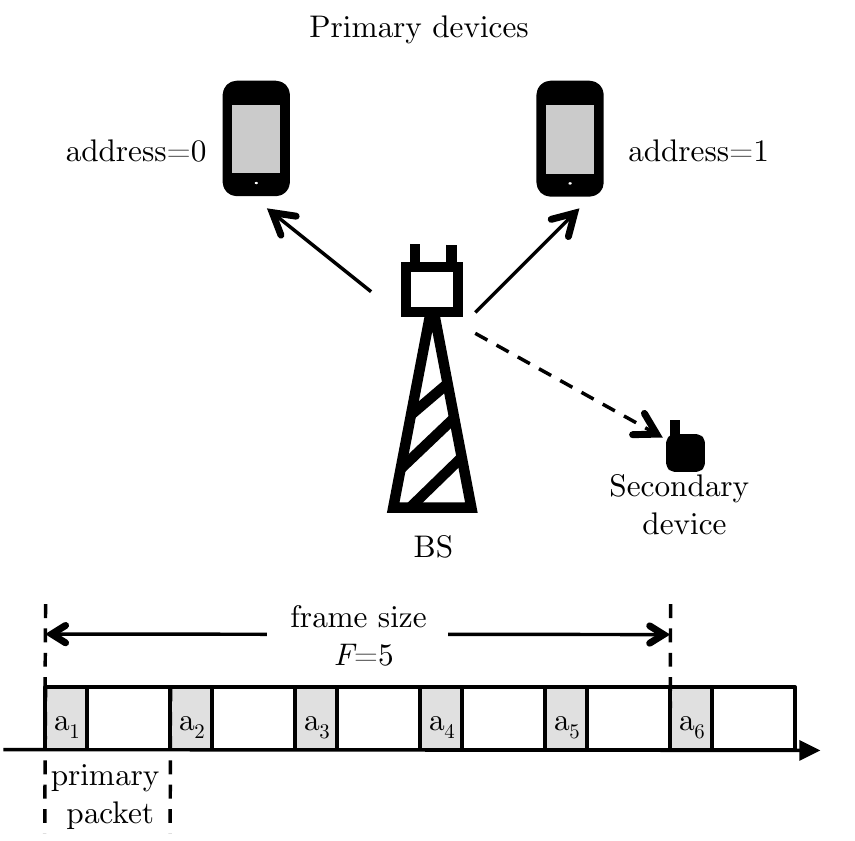}\\
  \caption{The primary system consists of a Base Station (BS) and two primary devices.
  Each primary packet has a header that contains address
  $a_i \in \{0,1\}$. The BS selects
  the orders of the packets in a frame in order to send
  information to the secondary device.}\label{fig:comm_scenario}
\end{figure}

The communication model is depicted on Fig.~\ref{fig:comm_scenario}.
A Base Station (BS)
transmits downlink data to a set of two users, addressed $0$ and
$1$, respectively. The BS serves the users in scheduling frames
with Time Division Multiple Access (TDMA). Each frame
has a fixed number of $F$ packets. Each packet carries the address
of a user to whom the packet is destined, as well as data for that
user. This is called \emph{primary} data, destined to either user
$0$ or user $1$. There is a third receiving device, termed \emph{secondary device},
that listens the TDMA frames sent by the BS. This device only
records the address of each packet and ignores
the packet data. Since this work is focused on the
secondary communication, the notions ``transmitter'' and ``receiver''
will be used to refer to secondary transmitter and receiver, respectively.
By addressing the packets in a
given frame in a particular order, the BS sends \emph{secondary
information}. Thus, an input symbol for
the secondary channel is an $F-$dimensional binary
vector $\mathbf{x} \in {\cal X}=\{0,1\}^F$.

The model with only two
primary is limiting, but extension to $K$ primary addresses
entails complexity that is outside the scope of this initial
paper on the topic. Yet, the results with binary secondary inputs
provide
novel insights for the communication strategies and set
the basis for generalizations to $K>2$.
Furthermore, the binary input captures the following
practical setup. Consider the case in which the arrival of packets
in the primary system is random and in a certain frame the BS has
only $F^{\prime}<F$ packets to send, then $(F-F^{\prime})$ of the
slots will be empty. In this case we can still use the binary input model. We assign address $0$ to a
the empty packet slots, such that these empty slots can be actually treated as 
valid secondary input symbols. On the other hand,
the presence of a packet in a given slot is treated as a secondary symbol $1$.
The secondary receiver
only needs to detect packet presence/absence, without
decoding its header.

The key assumption in the model is that the packets that are
scheduled in a frame are decided by the primary communication
system: the primary system decides that $s$
packets in a frame will be addressed to user $1$ and $(F-s)$
packets will be addressed to user $0$, where $0 \leq s \leq F$.
This assumption captures the essence of protocol coding: secondary
communication is realized by modulating the degrees of freedom
left over from the operation of the original, primary
communication system. In other words, it is assumed that the
operational requirements of the primary system are contained in
the set of packets that the BS decides to send in a given frame.

The number of packets $s$ addressed to user $1$ in a given frame
is called \emph{state} of the frame. We assume that the primary
system selects packets in a memoryless fashion: in each frame, a packet is addressed $1 (0)$ with
probability $a (1-a$), independently of
the other packets and the previous frames. Hence, the probability
that a frame is in state $s$ is binomial 
$P_S(s)=\binom{F}{s} a^s(1-a)^{F-s}$.
With the state  $s$  decided by the primary system,
the secondary transmitter is only allowed to rearrange
the packets in the frame. Since $s$ is a random variable
over which the secondary transmitter has no control, 
a frame carries a variable amount of secondary information. For example, if
$F=4$ and the primary system decides $s=3$, then the possible secondary
symbols for the frame are $1110, 1101, 1011, 0111$. But, if
$s=F=4$, than in that frame the secondary transmitter cannot send
any information.

\subsection{Error Models for the Secondary Channel}
\label{sec:ErrorModels}

From the
perspective of a secondary transmitter/receiver, each packet is
sent over a memoryless channel with binary inputs. Several suitable error models
can be inferred from the physical setup. In \emph{erasure channel}, the receiver either correctly decodes the packet address $0$ or $1$ or the
header checksum in incorrect, leading to erasure $\epsilon$.
In a \emph{binary symmetric channel}, the receiver uses error-correction decoding to decide whether it is more likely that address $0$ or $1$
is received. This results in only two possible outputs and symmetric error events.
Finally, the \emph{Z-channel} is suitable if $0/1$ corresponds to packet absence/presence, respectively.
The probability that, in absence of a packet, the noise produces a valid packet detection sequence, is practically $0$, while the probability
that packet transmission is not detected is $p_e>0$. 

In the general case of a channel with binary inputs, there can be $J$ possible
outputs from the set ${\cal J}$. The special cases above have ${\cal J}=\{0,1, \epsilon\}$ and ${\cal J}=\{0,1\}$. When $i=0,1$ is sent, there
are $J$ transition
probabilities, represented by a vector:
\begin{equation}\label{eq:Def_q_i}
\mathbf{q}_i=(q_{i1}, q_{i2}, \ldots q_{iJ}) \qquad i=0,1
\end{equation}
where $q_{ij}=P(y=j|x=i)$ and some $q_{ij}$ can be equal to $0$. A secondary output symbol is $\mathbf{y}
\in {\cal Y}={\cal J}^F$. The input/output variables of the
secondary channel are denoted by $\mathbf{X}$ and $\mathbf{Y}$,
respectively. By denoting $\mathbf{x} = (x_1, x_2, \cdots x_F )$ with $x_f \in \{0,1\} $ and $\mathbf{y} = (y_1, y_2, \cdots y_F )$ with $y_f \in {\cal J}$, we can define the channel $\mathbf{X}-\mathbf{Y}$ through the
transition probabilities:
\begin{equation}\label{eq:}
    P_{\mathbf{Y}|\mathbf{X}}(\mathbf{y}|\mathbf{x})=\prod_{f=1}^F
    q_{x_f y_f}
\end{equation}
When there is no risk for confusion, we
simply write $P(\mathbf{y}|\mathbf{x})$. Thus, the channel $\mathbf{X}-\mathbf{Y}$ is
specified by the memoryless binary
channel through which each packet is passed.

The following notation will be used. ${\cal S}=\{0, 1, \ldots F\}$
to denote the set of possible states. The set of input and output
symbols of the secondary channel is denoted by ${\cal X}$ and
${\cal Y}$, respectively. The set of input symbols is partitioned
into $F+1$ subsets ${\cal X}_s$ defined as follows:
\begin{equation}\label{eq:PartitionX}
\mathbf{x}  \in {\cal X}_s \Leftrightarrow \sum_{i=1}^Fx_i = s
\end{equation}
When the frame state is $S=s$, then only $\mathbf{x}  \in {\cal X}_s$ can be sent over the secondary
channel.


\section{Framework for Analyzing the Capacity of a Secondary Channel}


\subsection{Relation to the Shannon's Model
with Causal State Information at the Transmitter (CSIT)}

The secondary channel can be represented by the
framework of Shannon for channels
with causal state information at the transmitter (CSIT) \cite{ref:ShannonCSIT},. Shannon showed that instead of
considering the original channel with CSIT, one can consider an
ordinary, discrete memoryless channel with equivalent capacity
that has a larger input alphabet. The input variable of the
equivalent channel is $T$ and each possible input
letter $t$, termed \emph{strategy}~\cite{ref:CSITmonograph},
represents a mapping from the state alphabet ${\cal S}$ to the
input alphabet ${\cal X}$ of the original channel. A
particular strategy $t \in {\cal T}$ is defined by the vector of
size $|{\cal S}|$: $(t(1), \ldots t(|{\cal S}|))$, where $t(s) \in {\cal X}$.
Therefore, if each  $s \in {\cal S}$
can be mapped map to any $\mathbf{x} \in {\cal X}$, then the total
number of possible strategies is $|{\cal X}|^{|{\cal
S}|}$ and therefore $|{\cal T}| \leq |{\cal X}|^{|{\cal S}|}$.
The capacity of the equivalent channel can be found as:
\begin{equation}\label{eq:CapacityEquivalentCSIT}
  C=\max_{P_T(\cdot)} I(T, \mathbf{Y})
\end{equation}
where $P_T(\cdot)$ is a probability distribution defined over the
set ${\cal T}$ which is independent of the state $S$. The
maximization is performed across all the joint distributions that
satisfy~\cite{ref:CSITmonograph}:
\begin{equation} \label{eq:ShannotCSITjointdistributions}
   P_{S,T,\mathbf{X},\mathbf{Y}}(s,t,\mathbf{x},\mathbf{y})=P_S(s) P_T(t) \delta(\mathbf{x},t(s)) P_{\mathbf{Y}|\mathbf{X},S} (\mathbf{y}|\mathbf{x},s)
\end{equation}
where $\delta(\mathbf{x},t(s))=1$ if $\mathbf{x}=t(s)$ and
$\delta(\mathbf{x},t(s))=0$ otherwise. Following the properties of
mutual information (\cite{ThomasCover}, Section 8.3), the
required cardinality of ${\cal T}$ is not more than $|{\cal Y}|$.

However, Shannon's result is for the general case of channels
with causal CSIT. The secondary channel
considered here has a specific structure that permits more explicit
characterization of the
communication strategies. As noted in relation to
(\ref{eq:PartitionX}), for a given state $S=s$ only a subset
${\cal X}_s \in {\cal X}$ of symbols $\mathbf{x}$ may be produced.
For example, when $F=4$ and $s=2$, it is not possible
to send the symbol $\mathbf{x}=1011$. Nevertheless, in the model
with causal CSIT the distribution $P_{\mathbf{Y}|\mathbf{X},S}
(\mathbf{y}|\mathbf{x},s)$ needs to be defined for \emph{all
pairs} $(\mathbf{x}, s)$, irrespective of the fact that in the
original model some $\mathbf{x}$ are incompatible with $s$, i.~e.
when the state is $S=s$, the symbols $\mathbf{x} \notin {\cal
X}_s$ cannot be sent. In order to deal with this situation, we
need to extend the model. Given $P_{\mathbf{Y}|\mathbf{X}}(\mathbf{y}|\mathbf{x})$, we 
define $P_{\mathbf{Y}|\mathbf{X},S}(\mathbf{y}|\mathbf{x},s)$ in
the following way: For each $\mathbf{x}_u \notin {\cal X}_s$ we
take one $\mathbf{x}_v \in {\cal X}_s$ and define:
\begin{equation}
P_{\mathbf{Y}|\mathbf{X},S}(\mathbf{y}|\mathbf{x}_u,s) \equiv
P_{\mathbf{Y}|\mathbf{X},S}(\mathbf{y}|\mathbf{x}_v,s) \qquad
\forall \mathbf{y} \in {\cal Y}.
\end{equation}
The idea behind this approach is the following. For example, let us assume $F=4$ and the erasure model.
When $s=0$ only $\mathbf{x}=0000$ can be sent. But
we can look at it in another way: when $s=0$ only
$\mathbf{y}=0000$ ore the versions of $0000$ with erasures can occur. Hence,
we can equivalently say that when $s=0$, any $\mathbf{x}$ can be sent,
but, in absence of errors, the output is always $0000$.
Picking a strategy $t^{\prime \prime}$ in
which $t^{\prime \prime}(s)=\mathbf{x}_u$ is equivalent to picking
$t^{\prime}$ in which $t^{\prime}(s)=\mathbf{x}_v$.
In short, for given $s$, we define $P_{\mathbf{Y}|\mathbf{X},S}$ in order to discourage selection of symbols $\mathbf{x}$ for
which $\mathbf{x} \neq \mathbf{y}$ in absence of channel errors.


As pointed out in~\cite{ref:CSITmonograph}, expressing the
capacity in terms of strategies might pose some conceptual and
practical problems for code construction and implementation when
$F$ is large. On the other hand, our objective is to use the
specific way in which the set of states partitions the possible
set of transmitted symbols ${\cal X}$ in order to provide insights
in the capacity--achieving communication strategies. Therefore, a
different framework for capacity analysis from will be used. A practical dividend
of such a framework is presented in the companion paper~\cite{ref:Part2},
where the capacity--achieving strategies
are converted into convolutional code designs.

\subsection{Capacity Analysis through a Cascade of Channels}

Recall that $T$ is an auxiliary random variable defined over the
set of possible strategies ${\cal T}$. For given $T=t$ and
each $s \in {\cal S}$ there is a 
single \emph{representative of $t$ in $s$} $\mathbf{x}=t(s) \in {\cal
X}_s$. In the text that follows we use
``strategies'' and ``input symbols'' interchangeably. Hence, ${\cal T}$ consists of the
input symbols $\{ 1, 2, \ldots |{\cal T}| \}$. The set of $F+1$
representatives $\{ \mathbf{x}_s(t) \}$ for given $t$ will be
called a \emph{multisymbol} of $t$.

Due to the randomized state change, each $t \in {\cal T}$
induces a distribution on ${\cal X}$. For example, if $F=2$ and
the strategy is defined as $t(0)=00, t(1)=01, t(2)=11$, then we
can define $P_{\mathbf{X}|T}(\mathbf{x}=00|t)=(1-a)^2=P_S(0)$,
$P_{\mathbf{X}|T}(\mathbf{x}=11|t)=a^2=P_S(2)$, $P_{\mathbf{X}|T}(\mathbf{x}=01|t)=2a(1-a)=P_S(1)$,
and $P_{\mathbf{X}|T}(\mathbf{x}=10|t)=0$. In general, 
$P_{\mathbf{X}|T}(\cdot)$ should satisfy that for each
$s \in {\cal S}$ there is a single $\mathbf{x} \in {\cal X}_s$
such that $P_{\mathbf{X}|T}(\mathbf{x}|t)=P_S(s)$. The
set of such distributions is:
\begin{equation} \label{eq:PX|T}
{\cal P}_{\mathbf{X}|T} =\left \{P_{\mathbf{X}|T}(\cdot) | \forall
t \in {\cal T},\forall s \in {\cal S}, \exists ! \mathbf{x} \in
{\cal X}_s \textrm{ such that
}P_{\mathbf{X}|T}(\mathbf{x}|t)=P_S(s)\right\}
\end{equation}
In this way, we do not need to explicitly consider state in the
capacity analysis, but instead we model the secondary
communication channel by using a cascade of two channels $T - \mathbf{X} - \mathbf{Y}$ and  
the primary constraints are reflected in the definition of
${\cal P}_{\mathbf{X}|T}$. In order to express the mutual information $I(T;\mathbf{Y})$, we
write 
$I(T, \mathbf{X};\mathbf{Y}) = I(T;\mathbf{Y}) + I(\mathbf{X};\mathbf{Y}|T)
                        = I(\mathbf{X};\mathbf{Y}) + I(T;\mathbf{Y}|\mathbf{X})$
Using the Markov property for the cascade
we get $I(T;\mathbf{Y}|\mathbf{X})=0$, which implies:
\begin{equation}\label{eq:MutualInfITY}
    I(T;\mathbf{Y})=I(\mathbf{X};\mathbf{Y})-I(\mathbf{X};\mathbf{Y}|T)
\end{equation}
Let ${\cal P}_T$ denote the set of all distributions $P_T(\cdot)$.
Our objective is to find the pair of distributions $\left (
P_T(\cdot), P_{\mathbf{X}|T}(\cdot) \right )$ that maximizes
$I(T;\mathbf{Y})$. Thus, the capacity of the secondary channel is:
\begin{equation}\label{eq:MaxMutualInfITY}
    C=\max_{P_T(\cdot), P_{\mathbf{X}|T}(\cdot)}I(T;\mathbf{Y})
\end{equation}
We will always that $P_{\mathbf{X}|T}(\cdot)
\in {\cal P}_{\mathbf{X}|T}$ always. The
expression~(\ref{eq:MaxMutualInfITY}) can be upper--bounded:
\begin{equation}\label{eq:MaxMutualInfITY_upperbound}
   C \leq \max_{P_T(\cdot),
   P_{\mathbf{X}|T}(\cdot)}I(\mathbf{X};\mathbf{Y})- \min_{P_T(\cdot),
   P_{\mathbf{X}|T}(\cdot)} I(\mathbf{X};\mathbf{Y}|T)
\end{equation}
where the equality is achieved if and only if there is a pair of
distributions $\left ( P_T(\cdot), P_{\mathbf{X}|T}(\cdot) \right
)$ that simultaneously attains the max/min in the
first/second term, respectively. We will
decompose the problem~(\ref{eq:MaxMutualInfITY}) into two
sub--problems, maximization of $I(\mathbf{X};\mathbf{Y})$ and
minimization of $I(\mathbf{X};\mathbf{Y}|T)$.

\begin{figure}
\centering
\includegraphics[width=7cm]{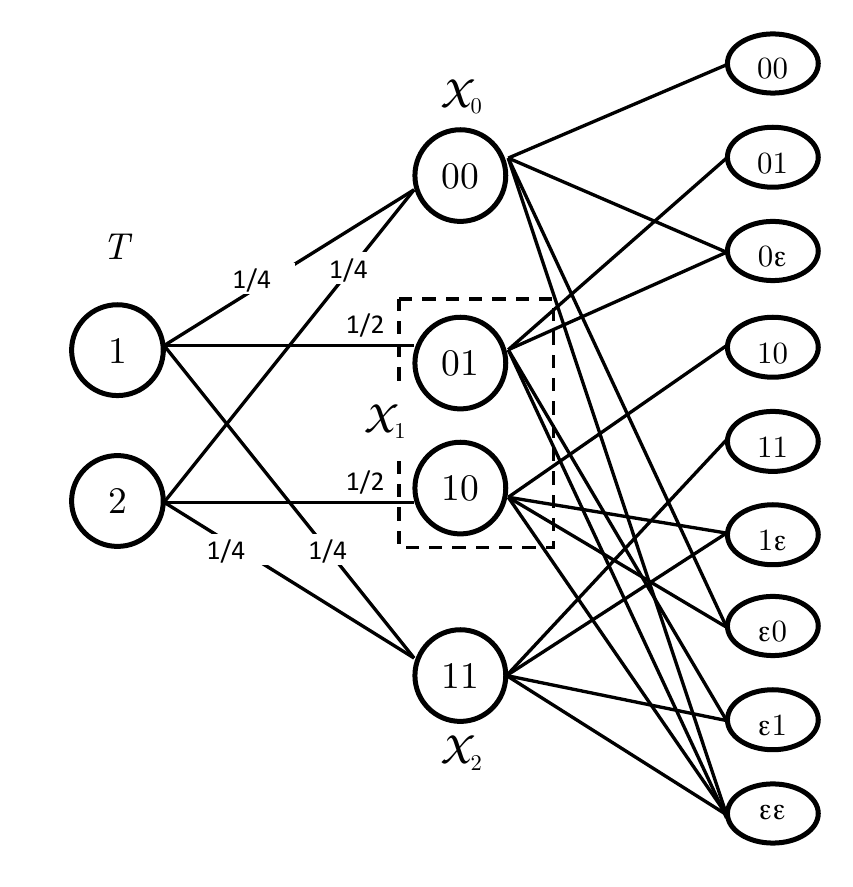}\\
\caption{Example choice of the probability distribution
$P_{\mathbf{X}|T}$ with $F=2$ and ${\cal T}=\{1,2\}$. The
transition probabilities on the channel $\mathbf{X}-\mathbf{Y}$
are not marked, but it is assumed that each packet $0$ or $1$ can
become erased $\epsilon$ independently with probability $p$.}
\label{fig:TransitionF2}
\end{figure}

Fig.~\ref{fig:TransitionF2} illustrates the cascade of channels where $F=2$ and erasure model for $\mathbf{X}-\mathbf{Y}$ with ${\cal J}=\{0,1,\epsilon \}$ and $q_{00}=q_{11}=1-p$,
while $q_{0 \epsilon}=q_{1 \epsilon}=p$. Let us assume that the
primary constraint uses $a=\frac{1}{2}$. The two multisymbols,
corresponding to $t=1$ and $t=2$ are $\{00, 01, 11\}$ and $\{00,
10, 11\}$, respectively. It is seen that uniform $P_T(\cdot)$ induces uniform $P_{\mathbf{X}}(\cdot)$. On
the other hand, the capacity of the vector channel with erasures
$\mathbf{X}-\mathbf{Y}$ is achieved when $P_{\mathbf{X}}(\cdot)$
is uniform. The reader can check that uniform
$P_T(\cdot)$ and the choice of $P_{\mathbf{X}|T}(\cdot)$ according
to Fig.~\ref{fig:TransitionF2}
simultaneously maximizes $I(\mathbf{X};\mathbf{Y})$ and minimizes
$I(\mathbf{X};\mathbf{Y}|T)$.

\section{Maximization of $I(\mathbf{X};\mathbf{Y})$}


Each pair of distributions $\left ( P_T(\cdot),
P_{\mathbf{X}|T}(\cdot) \right )$ induces a distribution
$P_{\mathbf{X}}$ on ${\cal X}$. Let ${\cal P}_{\mathbf{X}}$ denote
the set of all possible distributions $P_{\mathbf{X}}(\cdot)$,
while ${\cal P}^{T}_{\mathbf{X}} \subset {\cal P}_{\mathbf{X}}$
containing the distributions $P_{\mathbf{X}}(\cdot)$ that can be
induced by all possible pairs $\left ( P_T(\cdot),
P_{\mathbf{X}|T}(\cdot) \right )$. Then the following holds:
\begin{proposition}
The set of distributions ${\cal P}^{T}_{\mathbf{X}}$ is a subset
of $ {\cal P}_{\mathbf{X},S}$, where  $ {\cal P}_{\mathbf{X},S}
\subset {\cal P}_{\mathbf{X}}$ and:
\begin{equation} \label{eq:P_{X,S}}
{\cal P}_{\mathbf{X},S}=\left\{P_{\mathbf{X}}(\cdot) |
\sum_{\mathbf{x} \in {\cal X}_s} P_{\mathbf{X}}(\mathbf{x}) =
P_S(s), \forall s =0, 1, \cdots F \right \}
\end{equation}
\end{proposition}
\begin{IEEEproof}
We need to show that if $P_{\mathbf{X}}(\cdot) \in {\cal
P}^{T}_{\mathbf{X}}$, then $P_{\mathbf{X}}(\cdot) \in
{\cal P}_{\mathbf{X},S}$. Let $P_{\mathbf{X}}(\cdot) \in {\cal
P}^{T}_{\mathbf{X}}$, then:
\begin{equation}
  \sum_{\mathbf{x} \in {\cal X}_s} P_{\mathbf{X}}(\mathbf{x}) =
  \sum_{\mathbf{x} \in {\cal X}_s} \sum_{t \in {\cal T}} P_T(t) P_{\mathbf{X}|T}(\mathbf{x}|t) = \sum_{t \in {\cal T}}  P_T(t) \sum_{\mathbf{x} \in {\cal X}_s}
   P_{\mathbf{X}|T}(\mathbf{x}|t) \stackrel{(a)}{=}P_S(s)\sum_{t \in {\cal T}} P_T(t) \stackrel{(b)}{=} P_S(s) \nonumber
\end{equation}
where (a) follows from the definition~(\ref{eq:PX|T}) and (b) from $\sum_{t
\in {\cal T}} P_T(t)=1$.
\end{IEEEproof}

The previous proposition implies $\max_{P_T(\cdot), P_{\mathbf{X}|T}(\cdot)}I(\mathbf{X};\mathbf{Y})\leq \max_{P_{\mathbf{X}} (\cdot) \in {\cal P}_{\mathbf{X},S}}I(\mathbf{X};\mathbf{Y})$.
We will first look for the distribution $P_{\mathbf{X}^*}(\cdot)
\in {\cal P}_{\mathbf{X},S}$ that maximizes
$I(\mathbf{X};\mathbf{Y})$. Once $P_{\mathbf{X}^*}(\cdot)$ is known, we
choose $\left ( P_T(\cdot), P_{\mathbf{X}|T}(\cdot) \right )$ in order to induce the desired
$P_{\mathbf{X}^*}(\cdot)$. Let us define:
\begin{equation}\label{eq:DefinitionCXY}
    C_{XY}=\max_{P_{\mathbf{X}} \in {\cal P}_{\mathbf{X},S}(\cdot)} I(\mathbf{X};\mathbf{Y})
\end{equation}
which is never larger than the capacity of
$\mathbf{X}-\mathbf{Y}$, achieved by selecting over all $P_{\mathbf{X}}(\cdot) \in {\cal P}_{\mathbf{X}}$. For example, if the probability $a \neq \frac{1}{2}$ and there are erasure--type
errors, then $C_{XY}< F(1-p)$, where $F(1-p)$ is the capacity of $F$ erasure channel uses. This is because the achieving the capacity of the erasure channel requires uniform
distribution $P_{U, \mathbf{X}}(\mathbf{x})=2^{-F}$, which induces
the necessary condition 
$\sum_{\mathbf{x} \in {\cal X}_s} P_{U,\mathbf{X}}(\mathbf{x}) =
\binom{F}{s} 2^{-F}$, but this is not equal to $P_S(s)$ if $a \neq
\frac{1}{2}$.

In this text we are interested in channels $\mathbf{X}-\mathbf{Y}$
where each single channel use
$\mathbf{x}$ consists of $F$ uses of a more elementary, identical
channels, leading to the following symmetry: the set
of transition probabilities
$\{P_{\mathbf{Y}|\mathbf{X}}(\mathbf{y}|\mathbf{x})\}$ is
identical for all $\mathbf{x} \in {\cal X}_s$, as they are all
permutations of a vector with $s$ $1$s and $F-s$ $0$s. This is valid
irrespective of the the type of elementary channel used for a single
primary packet. Such a symmetry is
instrumental for making statements about $C_{XY}$. The following lemma is proved
in Appendix~\ref{app:Proof_Lemma_UniformInXS}.

\begin{lemma}
\label{lemma:UniformInXS}
The distribution $P_{\mathbf{X}}(\cdot) \in {\cal P_{X,S}}$ that
achieves $C_{XY}$ is, for all $s$ and each $\mathbf{x} \in {\cal
X}_s$:
\begin{equation} \label{eq:Condition_C_XY}
P_{\mathbf{X}}(\mathbf{x})=\frac{P_S(s)}{\binom{F}{s}}
\end{equation}
\end{lemma}


Having found $P_{\mathbf{X}}(\cdot)$ that attains $C_{XY}$,
it remains to find ${\cal T}$, $P_T(\cdot)$ and $P_{\mathbf{X}|T}(\cdot)$ (i. e. the representatives of each $T=t$)
such that~(\ref{eq:Condition_C_XY}) is satisfied. For example, let  $F=4$ and
$|{\cal X}_s|=1,4,6,4,1$ for $s=0,1,2,3,4$, respectively. Let at first take $|{\cal T}|=4m$ and uniform $P_T(t)=\frac{1}{4m}$. Then each $\mathbf{x} \in {\cal X}_1$ can be a representative of exactly $m$ different elements of ${\cal T}$, such that
$P_{\mathbf{X}}(\mathbf{X}=\mathbf{x})=P_S(1) \cdot m \cdot \frac{1}{4m}=P_S(1) / \binom{4}{1}$. In general, if $|{\cal T}|=\binom{F}{s} \cdot m$ and uniform $P_T(t)$, we can choose $\mathbf{x} \in {\cal X}_s$ to be a representative of exactly $m$ elements from ${\cal T}$; i.~e. $P_{\mathbf{X}|T}(\mathbf{x}|t)=P_S(s)$ for $m$ different values $t$ and zero otherwise. The resulting $P_{\mathbf{X}}(\cdot)$ satisfies~\eqref{eq:Condition_C_XY}. To satisfy this condition for all $s$ simultaneously, $|{\cal T}|$ should be divisible with $\binom{F}{s}$ for all $s=0 \cdots F$, leading to the following lemma, stated without proof (lcm stands for ``least common multiplier''):
\begin{lemma}
\label{lemma:lcm}
The distribution $P_{\mathbf{X}}(\cdot)$ that satisfies~\eqref{eq:Condition_C_XY} can be achieved by choosing uniform $P_T(\cdot)$ over a set with a minimal cardinality of $|\mathcal{T}|=\textrm{lcm}\left(\binom{F}{0},\binom{F}{1},\ldots,\binom{F}{F}\right)$.
\end{lemma}

\section{Minimization of $I(\mathbf{X};\mathbf{Y}|T)$}

\subsection{Definition of Minimal Multisymbols}
\label{sec:DefMinimalMultisymbols}

The multisymbol ${\cal M}_t=\{\mathbf{x}_0(t), \cdots \mathbf{x}_F (t)\}$ corresponding to $t$
has one representative in each $\mathbf{x}_s (t)={\cal X}_s$, such
that $P_{\mathbf{X}|T}(\mathbf{x}_s(t)|t)=P_S(s)$ and is zero for the other $\mathbf{x}$.
Since $I(\mathbf{X};\mathbf{Y}|T=t)$ depends on the choice of
representatives in ${\cal M}_t$, we will denote it by
$I(\mathbf{X};\mathbf{Y}|{\cal M}_t)$, such that:
\begin{equation}
I(\mathbf{X};\mathbf{Y}|T)=\sum_{t \in {\cal T}} I(\mathbf{X};\mathbf{Y}|{\cal M}_t)
\end{equation}
For example, let $F=5$ with
${\cal M}_1 = \{00000,00001,00011,00111,01111,11111\}$ and ${\cal M}_2 = \{00000,00001,00110,11100,10111,11111\}$. Assuming a binary symmetric channel with $q_{00}=q_{11}=0.8,
q_{01}=q_{10}=0.2$ it can be seen that $I(\mathbf{X};\mathbf{Y}|{\cal M}_1) < I(\mathbf{X};\mathbf{Y}|{\cal M}_2)$.
For intuitive explanation, consider two representatives $\mathbf{x}_{s_i} \in {\cal X}_{s_i}$,
$i=1,2$. From \eqref{eq:PartitionX} the Hamming weight of $\mathbf{x}_{s_i}$ is $s_i$ and,
without loss of generality, assume $s_1>s_2$. For the multisymbol
${\cal M}_1$, the Hamming distance between any
two representatives is given by:
\begin{equation}\label{eq:HammingDistances1s2}
    d_H(\mathbf{x}_{s_1},\mathbf{x}_{s_2})=s_2-s_1
\end{equation}
and is minimal possible. Informally, any two
representatives from ${\cal M}_1$ are as similar to each other as
possible since they represent the same
input $T=1$, which is not the
case for ${\cal M}_2$.

The multisymbols satisfying~(\ref{eq:HammingDistances1s2}) are of special interest
and will be termed \emph{minimal} multisymbols. Among them, there is one termed \emph{basic multisymbol} ${\cal M}^b$ with a particular structure: the representative in ${\cal X}_s$ is $00 \cdots 011 \cdots 1$  starts with $F-s$ consecutive zeros and $s$ consecutive ones. 
It can be shown that any minimal multisymbol can be obtained from the basic one
via permutation, such that there are $F!$ different minimal multisymbols. For example, let ${\cal M}^b=\{000, 001, 011, 111 \}$ and we apply the permutation $\pi=321$:
the components of each $\mathbf{x} \in {\cal M}^b$ are permuted according to $\pi$ to obtain ${\cal M}^m=\{000,
100, 110, 111 \}$. In general, for a given permutation $\pi$ we
define $\gamma_{\pi} (\cdot)$:
\begin{equation}\label{eq:PermutationMultisymbol}
    {\cal M}^{\prime}=\gamma_{\pi} ({\cal M})
\end{equation}
such that each $\mathbf{x}^{\prime}_s \in {\cal M}^{\prime}$ is obtained from the
corresponding $\mathbf{x}_s \in {\cal M}$ by permuting the packets according to
$\pi$ and the Hamming distance between any
two representatives is preserved $d_H(\mathbf{x}_{s_1},\mathbf{x}_{s_2})=
d_H(\mathbf{x}^{\prime}_{s_1},\mathbf{x}^{\prime}_{s_2})=s_2-s_1$.

\subsection{Analysis of $I(\mathbf{X};\mathbf{Y}|T=t)=I(\mathbf{X};\mathbf{Y}|{\cal M}_t)$}

We write the mutual information $I(\mathbf{X};\mathbf{Y}|{\cal M}_t)=H(\mathbf{Y}|{\cal M}_t)-H(\mathbf{Y}|\mathbf{X},{\cal M}_t)$ and first consider:
\begin{equation} \label{eq:HY|X_M_t}
H(\mathbf{Y}|\mathbf{X},{\cal M}_t)=\sum_{s=0}^{F} P_S(s) H(\mathbf{Y}|\mathbf{x}_s(t))
\end{equation}
Since each component of $\mathbf{x}_s$ uses identical memoryless channel, $H(\mathbf{Y}|\mathbf{x}_s(t))$
depends only on the Hamming weight $s$, but not on how the $0$s and $1$s are arranged in $\mathbf{x}_s$. This is stated through:
\begin{lemma}
\label{lemma3}
The conditional entropy for $\mathbf{x}_s \in {\cal X}_s$, having a Hamming weight of $s$, is given by:
\begin{equation}\label{eq:condi_entropy_HW_relation}
H(\mathbf{Y}|\mathbf{X}=\mathbf{x}_s)=sH(\mathbf{q}_1)+(F-s)
H(\mathbf{q}_0)=H_s
\end{equation}
where $H(\mathbf{q}_i)=-\sum_{j=1}^J q_{ij} \log_2 q_{ij}$ for $i=0,1$ and $\mathbf{q}_i$ is given by~(\ref{eq:Def_q_i}).
\end{lemma}
\begin{IEEEproof}
In order to determine $H(\mathbf{Y}|\mathbf{X}=\mathbf{x})=-\sum_{\mathbf{y} \in {\cal
J}^F} P(\mathbf{y}|\mathbf{x}) \log_2 P(\mathbf{y}|\mathbf{x})$, we use the fact that $P(\mathbf{y}|\mathbf{x})=\prod_{f=1}^F q_{x_f y_f}$ is a product distribution, such that we can write $H(\mathbf{Y}|\mathbf{X}=\mathbf{x})$ as:
\begin{eqnarray}
-\sum_{\mathbf{y} \in {\cal J}^F} \prod_{i=1}^F q_{x_iy_i} \sum_{j=1}^F \log_2 q_{x_jy_j} = - \sum_{j=1}^F \sum_{y_1 \in {\cal J}} \cdots \sum_{y_F \in {\cal J}} \log_2 q_{x_jy_j} \prod_{i=1}^F q_{x_iy_i} \nonumber
\end{eqnarray}
where (a) follows from changing the order of summation. If we consider the component $j=1$:
\begin{eqnarray}
- \sum_{y_1 \in {\cal J}} \cdots \sum_{y_F \in {\cal J}} \log_2 q_{x_1y_1} \prod_{i=2}^F q_{x_iy_i} &=&
-  \sum_{y_1 \in {\cal J}} q_{x_1y_1} \log_2 q_{x_1y_1} \sum_{y_2 \in {\cal J}} \cdots \sum_{y_F \in {\cal J}} \prod_{i=2}^F q_{x_iy_i} \nonumber \\
&\stackrel{(b)}{=}& -  \sum_{y_1 \in {\cal J}} \log_2 q_{x_1y_1} \cdot q_{x_1y_1} = H(\mathbf{q}_{x_1})
\end{eqnarray}
where (b) follows from $\sum_{y_2 \in {\cal J}} \cdots \sum_{y_F \in {\cal J}} \prod_{i=2}^F q_{x_iy_i}=1$. Doing the same for $j=2 \ldots F$ shows that
each $x_j=i$, $i=0,1$, contributes $H(\mathbf{q}_i)$ to
$H(\mathbf{Y}|\mathbf{X}=\mathbf{x})$, which proves the lemma.
\end{IEEEproof}
Using the lemma, \eqref{eq:HY|X_M_t} can be rewritten as $H(\mathbf{Y}|\mathbf{X},{\cal M}_t) = \sum_{s=0}^{F} P_S(s) H_s$
and is not affected
by the actual choice of ${\cal M}_t$, as long as there is a
representative in each ${\cal X}_s$.

\subsection{Analysis of $H(\mathbf{Y}|{\cal M}_t)$}

To gain intuition,
we first consider a special type of $P_S (\cdot)$, in which only two
states $s_1, s_2 \in {\cal S}$ occur with non-zero
probability $P_S(s_1)=\lambda$ and
$P_S(s_2)=1-\lambda$, such that ${\cal
M}_t=\{ \mathbf{x}_{s_1}, \mathbf{x}_{s_2} \}$. Due to the symmetry
implied by Lemma~\ref{lemma3}, without losing generality, we
first pick an arbitrary $\mathbf{x}_{s_1} \in {\cal
X}_{s_1}$. Then, how to select $\mathbf{x}_{s_2} \in {\cal X}_{s_2}$ in order to minimize the $H(\mathbf{Y}|{\cal M}_t)$? Slightly abusing the notation from \eqref{eq:HammingDistances1s2}, we use $d_H(\mathbf{x})$ to denote the Hamming weight of
$\mathbf{x}$. Recall that $d_H(\mathbf{x})=s$ for $\mathbf{x} \in {\cal X}_s$. Let
$g_{uv}(\mathbf{x}_{s_1},\mathbf{x}_{s_2})$, where $u, v
\in\{0,1\}$ denote the number of positions $f$ at which
$x_{s_1f}=u$ and $x_{s_2f}=v$. For example, if
$\mathbf{x}_{s_1}=00110$, $\mathbf{x}_{s_2}=11011$, then
$g_{00}=0$,  $g_{01}=3$,  $g_{10}=1$, and $g_{11}=1$ (we write $g_{uv}$ for brevity).
Using similar arithmetics as in Lemma~\ref{lemma3}:
\begin{eqnarray} \label{eq:HYgivenT_with_two_inputs}
H(\mathbf{Y}|T=t) \! = \! g_{00}H(\mathbf{q}_0)+g_{11}H(\mathbf{q}_1)+
g_{01}H(\lambda \mathbf{q}_0 + (1\textrm{-}\lambda) \mathbf{q}_1) +
g_{10}H((1\textrm{-}\lambda) \mathbf{q}_0 + \lambda \mathbf{q}_1)
\end{eqnarray}
The Hamming distance is
$d_H(\mathbf{x}_{s_1},\mathbf{x}_{s_2})=g_{01}+g_{10}$. 
The following lemma formalizes the intuition that $H(\mathbf{Y}|{\cal
M}_t)$ is minimized when any two representatives are as similar to
each other as possible.
\begin{lemma}
When ${\cal M}_t$ consists of only two
representatives $\mathbf{x}_{s_1},\mathbf{x}_{s_2}$, $H(\mathbf{Y}|{\cal
M}_t)$ is minimized when the Hamming distance
$d_H(\mathbf{x}_{s_1},\mathbf{x}_{s_2})=|s_2-s_1|$ is minimal possible.
\end{lemma}
\begin{IEEEproof}
Without loss of generality, assume that $s_2>s_1$. Then
    $g_{10}(\mathbf{x}_{s_1},\mathbf{x}_{s_2})<g_{01}(\mathbf{x}_{s_1},\mathbf{x}_{s_2})$
since $d_H(\mathbf{x}_{s_1})<d_H(\mathbf{x}_{s_2})$. Assume that
$g_{10}(\mathbf{x}_{s_1},\mathbf{x}_{s_2})>0$ and let there be $f_1, f_2$ such that:
\begin{eqnarray}\label{eq:Lemma_g10_g01_2}
    (x_{s_1,f_1},x_{s_2,f_1})=(1,0) \qquad
    (x_{s_1,f_2},x_{s_2,f_2})=(0,1)
\end{eqnarray}
Let $\mathbf{z}_{s_2}$ be another representative from ${\cal
X}_{s_2}$,  obtained by swapping the positions $f_1,f_2$
in $\mathbf{x}_{s_2}$, but keeping the other values of
$\mathbf{x}_{s_2}$, such that $z_{s_2,f_1}=1$ and $z_{s_2,f_2}=0$.
Then:
\begin{eqnarray}\label{eq:Lemma_g10_g01_3}
  g_{00}(\mathbf{x}_{s_1},\mathbf{x}_{s_2})+1= g_{00}(\mathbf{z}_{s_1},\mathbf{z}_{s_2}) \qquad
  g_{11}(\mathbf{x}_{s_1},\mathbf{x}_{s_2})+1 =
  g_{11}(\mathbf{z}_{s_1},\mathbf{z}_{s_2}) \nonumber \\
  g_{01}(\mathbf{x}_{s_1},\mathbf{x}_{s_2})-1 =
  g_{01}(\mathbf{z}_{s_1},\mathbf{z}_{s_2}) \qquad
  g_{10}(\mathbf{x}_{s_1},\mathbf{x}_{s_2})-1 =
  g_{10}(\mathbf{z}_{s_1},\mathbf{z}_{s_2})
\end{eqnarray}
Using the concavity of the entropy function, we can write:
{\small
\begin{eqnarray}\label{eq:Lemma_g10_g01_4}
    H(\lambda \mathbf{q}_0 + (1\textrm{-}\lambda) \mathbf{q}_1)+H((1\textrm{-}\lambda)
\mathbf{q}_0 + \lambda \mathbf{q}_1) \geq
    \lambda H(\mathbf{q}_0) + (1\textrm{-}\lambda) H(\mathbf{q}_1)+
    (1\textrm{-}\lambda)H(\mathbf{q}_0) + \lambda H(\mathbf{q}_1)=
    H(\mathbf{q}_0)+H(\mathbf{q}_1)
\end{eqnarray}}
Using~\eqref{eq:Lemma_g10_g01_3} and~\eqref{eq:Lemma_g10_g01_4} it
follows:
{\small \begin{eqnarray}\label{eq:Lemma_g10_g01_4}
H_{\mathbf{x}_{s_1},\mathbf{x}_{s_2}}=
g_{00}H(\mathbf{q}_0) +
g_{11}H(\mathbf{q}_1) + g_{01}H(\lambda
\mathbf{q}_0 + (1-\lambda) \mathbf{q}_1)+
g_{10}H((1-\lambda)
\mathbf{q}_0 + \lambda \mathbf{q}_1)
\geq \nonumber \\
g_{00}H(\mathbf{q}_0) +
g_{11}H(\mathbf{q}_1) + (g_{01}-1)H(\lambda
\mathbf{q}_0 + (1-\lambda) \mathbf{q}_1)+
(g_{10}-1)H((1-\lambda)
\mathbf{q}_0 + \lambda \mathbf{q}_1)=
H_{\mathbf{x}_{s_1},\mathbf{z}_{s_2}}\nonumber
\end{eqnarray}}
\noindent where $g_{uv}=g_{uv}(\mathbf{x}_{s_1},\mathbf{x}_{s_2})$ and $H_{\mathbf{x}_{s_1},\mathbf{x}_{s_2}}= H(\mathbf{Y}|{\cal M}_t=\{\mathbf{x}_{s_1},\mathbf{x}_{s_2} \})$. We can analogously continue the swap the positions in
$\mathbf{x}_{s_2}$ until getting $g_{10}=0$. Each swap does not
increase $H(\mathbf{Y}|{\cal M}_t)$, which means that
when $g_{10}=0$, $H(\mathbf{Y}|{\cal M}_t)$ is minimal.
\end{IEEEproof}

We now consider a general $P_S(\cdot)$. As indicated above,
$H(\mathbf{Y}|{\cal M}_t)$ can be written as:
\begin{equation}\label{eq:entropy_position_members}
    H(\mathbf{Y}|{\cal M}_t)=\sum_{f=1}^F H(\mathbf{u}_f)
\end{equation}
where $\mathbf{u}_f$ is the probability distribution that
corresponds to the $f-$th position, defined as:
\begin{equation}\label{eq:distribution_u_f}
    \mathbf{u}_f=\sum_{s=0}^F P_s\left[(1-x_{s,f})\mathbf{q}_0 + x_{s,f}\mathbf{q}_1 \right] \qquad \textrm{where }x_{s,f} \in \{0,1\}
\end{equation}

Without losing generality, let us take the first value $x_{s1}$ of
each of the representatives $\mathbf{x}_s$ can create
$(F+1)-$dimensional vector $\mathbf{z}_1$. In a similar way
$\mathbf{z}_2$ is created, such that:
\begin{equation}\label{eq:vectors_z1_z2}
  \mathbf{z}_1 = (x_{01}, x_{11}, \cdots x_{F1}) \qquad
  \mathbf{z}_2 = (x_{02}, x_{12}, \cdots x_{F2})
\end{equation}
The probability distribution vectors $\mathbf{u}_1$ and
$\mathbf{u}_2$ can be written as:
\begin{eqnarray} \label{eq:u1_u2_distributions}
  \mathbf{u}_1 = (Q_{00}+Q_{01}) \mathbf{q}_0 + (Q_{10}+Q_{11}) \mathbf{q}_1 \qquad
  \mathbf{u}_2 = (Q_{00}+Q_{10}) \mathbf{q}_0 + (Q_{01}+Q_{11}) \mathbf{q}_1
\end{eqnarray}
where $Q_{uv}=\sum_{s \in {\cal G}_{uv}(z_1,z_2)} P_s$ and
the sets ${\cal G}_{uv}(z_1,z_2)=\{s | x_{s,1}=u, x_{s,2}=v\}$
for $u,v \in \{0,1\}$.

\begin{lemma} \label{lemma:Empty_Sets}
The contribution of the positions $1$ and $2$ to the entropy
$H(\mathbf{Y}|{\cal M}_t)$ is minimized when one of the sets
${\cal G}_{01}, {\cal G}_{10}$ is empty.
\end{lemma}
\begin{IEEEproof} Let us start with a multisymbol  $\{\mathbf{x}_s\}$
in which none of the sets ${\cal G}_{01}(z_1,z_2), {\cal
G}_{10}(z_1,z_2)$ is empty. Without losing generality, we will
``empty'' the set ${\cal G}_{01}(z_1,z_2)$ as follows: If there is $s \in {\cal S}$ such
that $x_{s,1}=0, x_{s,2}=1$, these two positions in the
representative $\mathbf{x}_s$ are swapped. That is, if there is a
representative $\mathbf{x}=01 \cdots$, it is changed to $10
\cdots$. Using the concavity of the entropy, we can show that
these swapping operations can decrease the contribution of the
positions $f=1,2$ to the entropy
\eqref{eq:entropy_position_members}. Note that after swapping
\eqref{eq:u1_u2_distributions}, the new distributions are:
\begin{eqnarray} \label{eq:u1_u2_new_distributions}
  \mathbf{u}^{\prime}_1 = Q_{00} \mathbf{q}_0 + (Q_{01}+Q_{10}+Q_{11}) \mathbf{q}_1 \qquad
  \mathbf{u}^{\prime}_2 = (Q_{00}+Q_{01}+Q_{10}) \mathbf{q}_0 + Q_{11} \mathbf{q}_1
\end{eqnarray}
Using the concavity property, it can be shown that  
\begin{equation}\label{eq:u1primeu2primecontributions}
    H(\mathbf{u}_1)+H(\mathbf{u}_2) \geq H(\mathbf{u}^{\prime}_1)+H(\mathbf{u}^{\prime}_2)
\end{equation}
where $\mathbf{u}_1, \mathbf{u}_2$ and $\mathbf{u}^{\prime}_1,
\mathbf{u}^{\prime}_2$ are given by \eqref{eq:u1_u2_distributions}
and \eqref{eq:u1_u2_new_distributions}, respectively. Analogously,
the contribution from the two positions
will decrease to the value
\eqref{eq:u1primeu2primecontributions} if the set ${\cal
G}_{10}(z_1,z_2)$ is emptied.
\end{IEEEproof}

This analysis leads us to the following theorem (proof in Appendix~\ref{sec:ProofTheoremMinimalMultisymbol}) and corollary:
\begin{theorem}
\label{TheoremMinimalMultisymbol}
When each individual packet in a frame is sent over an identical
channel with binary inputs and general outputs, the minimal
multisymbol minimizes $H(\mathbf{Y}|{\cal M}_t)$.
\end{theorem}
\begin{corollary}
The following mutual information is constant for all minimal multisymbols ${\cal M}^m$:
\begin{equation}\label{eq:Im}
I(\mathbf{X};\mathbf{Y}|{\cal M}^m)=H(\mathbf{Y}|{\cal M}^m)-H(\mathbf{Y}|\mathbf{X},{\cal M}^m)=I_m
\end{equation}
\end{corollary}

\section{Achieving the Capacity of the Secondary Channel}

Here we analyze \eqref{eq:MaxMutualInfITY_upperbound} and find ${\cal T}$ and $\{ {\cal M}_t \}$ (i.~e. $P_T(\cdot))$ and $P_{\mathbf{X}|T}(\cdot)$, respectively) that simultaneously maximizes $I(\mathbf{X};\mathbf{Y})$ according to Lemma~\ref{lemma:UniformInXS} and minimizes $I(\mathbf{X};\mathbf{Y}|T)=I_m$ according to \eqref{eq:Im}. Recall
that uniform $T$ with $|{\cal T}|=\textrm{lcm}\left(\binom{F}{0},\binom{F}{1},\ldots,\binom{F}{F}\right)=L$ can achieve $C_{XY}$. Since there are $F! \geq L$ multisymbols, then in principle it should be possible to select $L$ minimal multisymbols in order to have $I(\mathbf{X};\mathbf{Y}|T)=I_m$ and maximize $I(\mathbf{X};\mathbf{Y})$.

\begin{figure}
\begin{minipage}[b]{0.5\linewidth}
\centering
{ \footnotesize
\begin{tabular}{|c|c|}
\hline
$t$ & $ \{ \mathbf{x}_s(t) \}$ \\
\hline
$1$ & $(0000,0001,0011,0111,1111)$\\
\hline
$2$ & $(0000,0001,0101,0111,1111)$ \\
\hline
$3$ & $(0000,0001,1001,1011,1111)$ \\
\hline
$4$ & $(0000,0010,0011,0111,1111)$ \\
\hline
$5$ & $(0000,0010,0110,0111,1111)$ \\
\hline
$6$ & $(0000,0010,1010,1011,1111)$ \\
\hline
$7$ & $(0000,0100,0101,0111,1111)$ \\
\hline
$8$ & $(0000,0100,0110,0111,1111)$ \\
\hline
$9$ & $(0000,0100,1100,1101,1111)$ \\
\hline
$10$ & $(0000,1000,1001,1101,1111)$ \\
\hline
$11$ & $(0000,1000,1010,1110,1111)$ \\
\hline
$12$ & $(0000,1000,1100,1110,1111)$ \\
\hline
\end{tabular}
}

{\footnotesize (a)}
\label{fig:figure1}
\end{minipage}
\hspace{0.5cm}
\begin{minipage}[b]{0.5\linewidth}
\centering
\includegraphics[width=8cm]{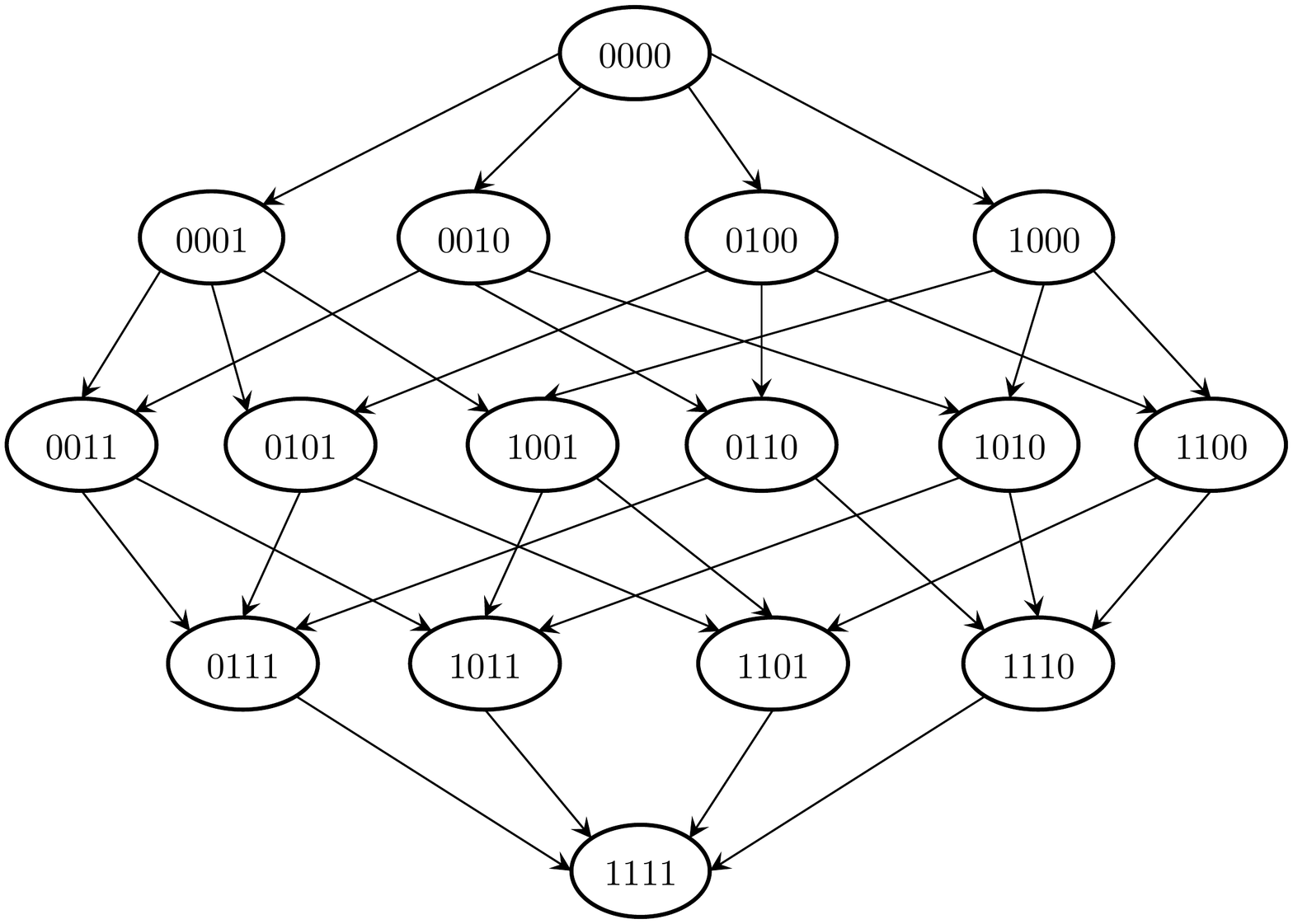}\\
{\footnotesize (b)}
\label{fig:figure2}
\end{minipage}
\caption{Selection of the representative sets for $F=4$ that achieve the capacity. (a) Multisymbols for the $12$ inputs (b) Graph representation of the process for selecting the multisymbols $\mathbf{x}_s(t)$.}
\label{fig:F4graph_layer}
\end{figure}

In order to show that it is always possible to select $\{ {\cal M}_t \}$, with  $|\{ {\cal M}_t \}|=L$ and uniform $T$, we first take an example with $F=4$. The set of $L=12$ multisymbols can be selected as on Fig.~\ref{fig:F4graph_layer}(a). Multisymbols can be represented by a directed graph, see Fig.~\ref{fig:F4graph_layer}(b). Each node in the graph represents a particular $\mathbf{x} \in {\cal X}$. An edge exists between $\mathbf{x}_s \in {\cal X}_s$ and $\mathbf{x}_{s+1} \in {\cal X}_{s+1}$ if and only if the Hamming distance is $d_H(\mathbf{x}_s,\mathbf{x}_{s+1})=1$. The directed edge from $\mathbf{x}_s$ to $\mathbf{x}_{s+1}$ exists if they can both belong to a same minimal multisymbol ${\cal M}_t$. A multisymbol is represented by a path of length $F$ that starts at $00 \cdots 0$ and ends at $11 \cdots 1$. To each edge we can assign a nonnegative integer, which denotes the number of multisymbols (paths) that contain that edge. On Fig.~\ref{fig:F4graph_layer}(b), each edge that starts from $0000$ has a weight $3$, each edge between an element of ${\cal X}_1$ and ${\cal X}_2$ has a weight $1$, etc. The weight of each edge between $\mathbf{x}_s$ and $\mathbf{x}_{s+1}$ can be treated as an outgoing weight for $\mathbf{x}_s$ and incoming weight for $\mathbf{x}_{s+1}$. Using this framework, we need to prove that, for each $s=0 \ldots F-1$, it is possible to match all outgoing weights from ${\cal X}_s$ to all incoming weights from ${\cal X}_{s+1}$. This is stated with the following theorem (proof in Appendix~\ref{sec:ProofTheorem_Possible_choice}):
\begin{theorem}
\label{Theorem_Possible_choice}
If $L=\textrm{lcm}\left(\binom{F}{0},\binom{F}{1},\ldots,\binom{F}{F}\right)$ and the distribution over ${\cal T}$ is uniform, then the multisymbols can be chosen such as to achieve the capacity of the secondary channel.
\end{theorem}

If $F=4$ it turns out that $\frac{m_s}{F-s}$ is always an integer, such that all the outgoing/incoming weights to the same node are identical. This is not the case if, e.~g., $F=7$, then $L=105$, $m_1=15$ and $\frac{m_1}{7-1}=\frac{15}{6}$, such that each node from ${\cal X}_1$ has $3$ outgoing edges of weight $3$ and $3$ of weight $2$.

%
%


\section{Further Considerations and Numerical Illustrations}


In absence of errors $\mathbf{Y}=\mathbf{X}$, such that $I(T;\mathbf{Y})=I(T;\mathbf{X})$ and the capacity is
\begin{equation}\label{eq:CapacityComErrorless}
    C_{F,0}=\sum_{s=0}^F P_S(s) \log_2 \binom{F}{s}
\end{equation}
When there are no errors, the state $s$ is always known also at the receiver and the communication strategy is different, see~\cite{PetarGC10}. Each state $s$ is seen as a different subchannel, also denoted $s$, and both the transmitter $\mathbf{X}$ and the receiver $\mathbf{Y}$ know which subchannel is used in a frame. Let $r(F,s)=\log_2 \binom{F}{s}$ denote the number of bits that are sent in a single use of the subchannel $s$. Considering a large number of channel uses $n \rightarrow \infty$, then the realization of the sequence of frame states becomes typical~\cite{ThomasCover} and the state $s$ occurs approximately $nP_S(s)$ times. The sender segments the message into submessages and each submessage is sent over a separate subchannel. The submessage sent over the
subchannel $s$ contains approximately $nP(s)r_(F,s)$ bits. If during the $i-$th channel use the sender observes that the state $s$, then
it takes the next $r(F,s)$ bits from the corresponding submessage. Thus, the whole message is sent by time--interleaving of all the available subchannels and the time--interleaved sequence is perfectly observed by the receiver.

\begin{figure}
  \centering
  \includegraphics[width=8.3cm]{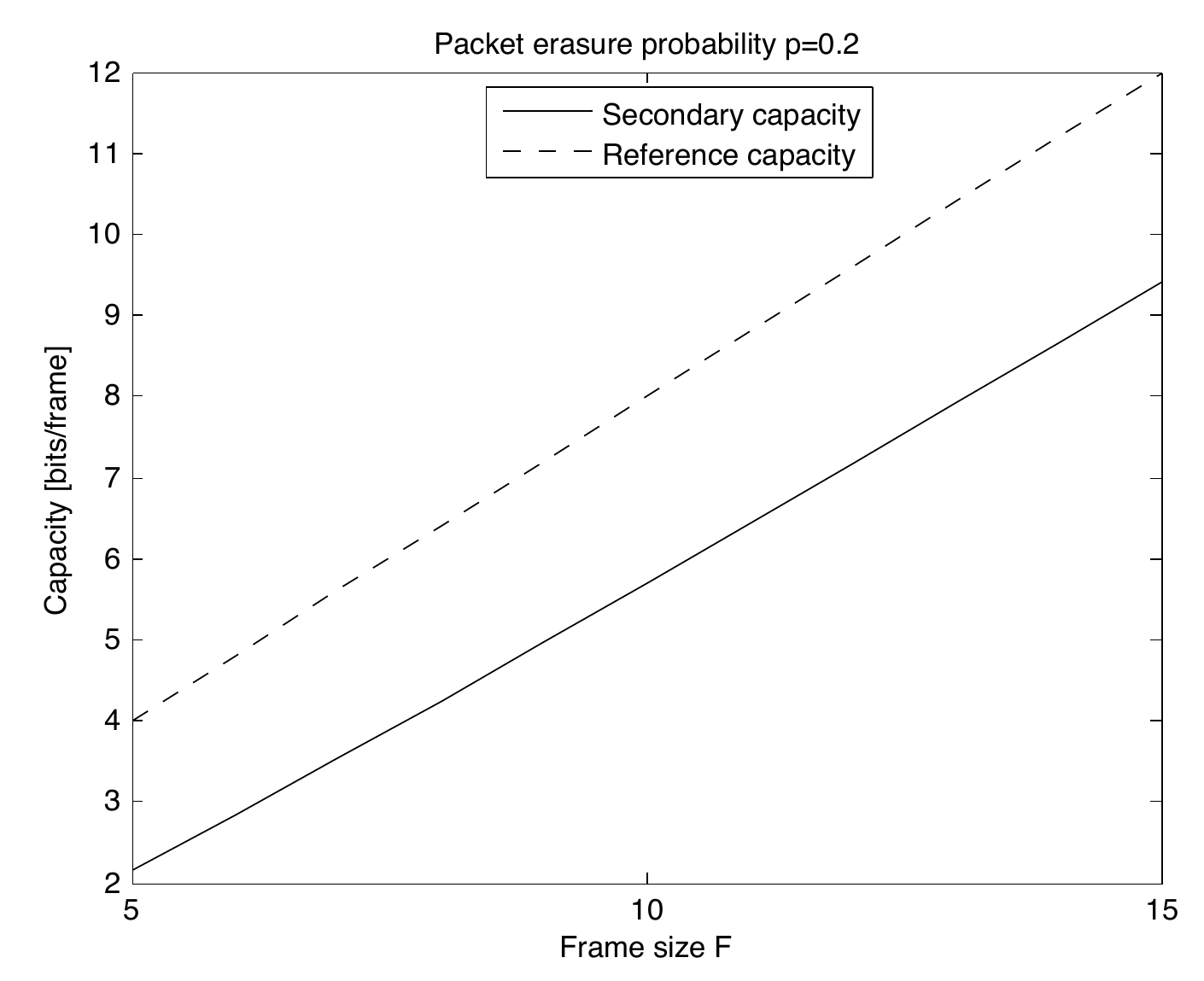}\\
  \caption{Comparison between the secondary capacity with the reference capacity (erasure channel with $F$ uses). The probability $a=\frac{1}{2}$}\label{fig:CvsF_p02_erasure}
\end{figure}

We now consider the model with erasures. An upper bound on the secondary capacity is simply taking $C_{XY}$, as defined in~(\ref{eq:DefinitionCXY}). If $a=\frac{1}{2}$, then $C_{XY}=F(1-p)$, the capacity of the erasure channel with $F$ uses. Consider now the asymptotic case $F \rightarrow \infty$ and observe a single frame (one single channel use). The state becomes typical and, with high probability,  $s \in \left (\frac{F(1- \epsilon)}{2}, \frac{F(1+ \epsilon)}{2} \right )$, where $\epsilon \rightarrow 0$ as $F \rightarrow \infty$. We sketch how the capacity can be achieved in this case.  First note that it suffices that the ${\cal T}$ is $\binom{F}{\frac{F(1- \epsilon)}{2}}$, where the latter is assumed to be integer. Then a multisymbol for each $T=t$ has representatives in the sets ${\cal X}_s$, where $s \in \left [\frac{F(1- \epsilon)}{2}, \frac{F(1+ \epsilon)}{2} \right ]$. If a state $s$ outside of that interval occurs, then an arbitrary $\mathbf{x}$ is sent. With this strategy, there are some $\mathbf{x} \in {\cal X}_s$ with $s>\frac{F(1- \epsilon)}{2}$ that are unused, but this is asymptotically negligible, and it can be shown that
\begin{equation} \label{eq:erasuregap}
\lim _{F \rightarrow \infty} \frac{C_F}{F}=(1-p)
\end{equation}
where $C_F$ is the capacity when the frame size is $F$. In other words, the normalized capacity approaches the capacity of a binary erasure channel, which is expected. A numerical illustration for the erasure channel is given on Fig.~\ref{fig:CvsF_p02_erasure} and it can be seen that for relatively small $F$, the gap between the capacity of the secondary channel and $F-$uses erasure channel is substantial. Note that the equation (\ref{eq:erasuregap}) does not state that the gap will disappear, but only that it is of a type $o(F)$, i. e. becomes asymptotically zero compared to $F$.

\begin{figure}
  \centering
  \includegraphics[width=14cm]{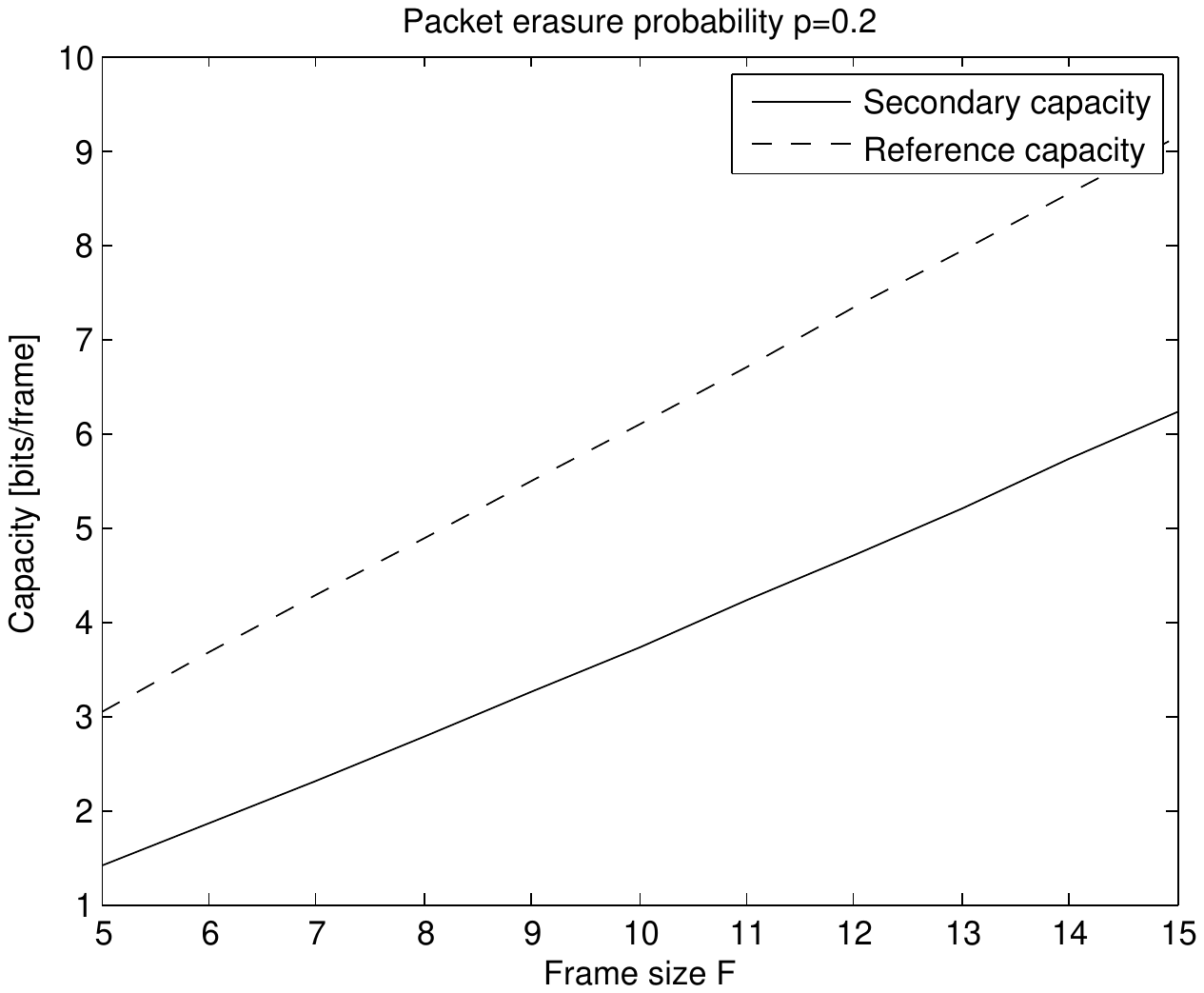}\\
  \caption{Comparison between the capacity of the combinatorial model and the outer bound for $p=0.2$ and $a=0.5$.}\label{fig:F_Z}
\end{figure}

We finally consider the case of a Z-channel, introduced in Section~\ref{sec:ErrorModels}. Recall that this is suitable 
when address $0$ is an ``empty'' user, while address $1$ means that
there is a packet transmission (irrespective to which user it is
addressed). The capacity of a binary $Z$-channel with crossover
probability $p$ is given by $
C_{Z}(p)=\log_2\left(1+(1-p)p^{p/(1-p)}\right)$.
The capacity--achieving distribution for the $Z-$channel
requires nonuniform input distribution $P_{U,
\mathbf{X}}(\mathbf{x})\neq 2^{-F}$. As a simple outer bound on the capacity of the secondary channel, we again take $C_{XY}$, which for given input probability $a$
is given by $C_{XY}=C_{F, out}=FC_{Z,a}(p)$, 
where $C_{Z,a}(p)$ is the capacity of the binary $Z$-channel under a fixed value of the input probability $a$, given by $C_{Z,a}(p)= -\left(1-a(1-p)\right)\log_2\left(1-a(1-p)+a(1-p)\log_2
a(1-p)\right)\nonumber +a\left(p\log_2(p)+(1-p)\log_2
(1-p)\right)$. Some illustrative results for the $Z-$channel modelare provided on Fig.~\ref{fig:F_Z}. The channel
capacity is compared to the outer bound
in dependency of the frame length $F$, for a fixed crossover probability $p=0.2$ and $a=0.5$.

Similar to the discussion for the erasure channel, for the $Z-$channel we
also consider the asymptotic case $F \rightarrow \infty$ and
observe a single frame (channel use). 
Using similar arguments as for the erasure channel, for the asymptotic case with a $Z-$channel model it can be shown that
\begin{equation}
\lim _{F \rightarrow \infty} \frac{C_F}{F}= C_{Z,a}(p)
\end{equation}

\section{Conclusion}

We have introduced a class of communication channels with
protocol coding, i.~e. the information is modulated in the actions
taken by the communication protocol of an existing, primary
system. In particular, we have considered strategies in which
protocol coding is done by combinatorial ordering of the labelled user
resources (packets, channels) in the primary system. Differently from the previous works, our focus here is not on the steganographic usage of this type of protocol coding. Our aim is rather on its ability to introduce a new 
\emph{secondary communication channel}, intended for reliable communication with newly introduced secondary devices, that are low-complexity versions of the primary devices, capable only to decode the robustly encoded header information in the primary signals. The key feature of the communication model is that it captures the constraints that the primary system operation puts on protocol coding i. e. the secondary information can only be sent by rearranging the set of packets made available by the primary system. The challenge is that the amount of information that can be sent in this way is not controllable by the secondary - e. g. if the all the primary packets in a given scheduling epoch carry the same label, then all re-arrangements look equivalent to a secondary receiver and no secondary information can be sent. Since the main application of the secondary channels introduced here is reliable communication, we have focused on investigating the communication strategies that can be used under various error models. We have derived the capacity of the secondary channel under arbitrary error models. The insights obtained from the capacity--achieving communication strategies are used in Part II of this work to design practical error--correcting mechanisms for secondary channels with protocol coding. 

\section*{Acknowledgment}
The authors would like to thank Prof. Osvaldo Simeone (New Jersey Institute of Technology) for useful discussions on the channels with causal channel state information at the transmitter.

\appendix

\subsection{Proof of Lemma~\ref{lemma:UniformInXS}}
\label{app:Proof_Lemma_UniformInXS}

\begin{proof}
We generalize the Theorem 4.5.1 from
\cite{ref:Gallager} to reflect the fact that the maximization is
over ${\cal P}_{\mathbf{X},S}$ rather than ${\cal P}_{\mathbf{X}}$.
Let us denote
$P_{\mathbf{X}}(\mathbf{x}_{s,k})=\alpha_{s,k}$ where
$\mathbf{x}_{s,k}$ is the $k-$th element (e. g. in a lexicographic
order) within the set ${\cal X}_s$. Let ${\boldsymbol
\alpha}=(\alpha_{0,1}, \alpha_{1,1}, \alpha_{1,2}, \ldots,
\alpha_{F,F})$ be the $2^F$-dimensional probability vector. Then
$I(\mathbf{X},\mathbf{Y})=f(\mathbf{\alpha})$ and the maximization
problem is:
\begin{eqnarray}
\max f(\mathbf{{\boldsymbol \alpha}}) \qquad \textrm{ such that }
\quad \sum_{k=1}^{K_s} \alpha_{s,k}=p_s, \quad \forall s \in {\cal S}
\end{eqnarray}
where $p_s=P_S(s)$ and $K_s=|{\cal
X}_s|=\binom{F}{s}$. The constraint $\sum_{s,k}
\alpha_{s,k}=1$ is redundant, since $\sum_{s} p_s=1$. We need to
use $(F+1)$ Lagrangian multipliers and maximize $f({\boldsymbol
\alpha})-\sum_s \lambda_s (\sum_{k} \alpha_{s,k} - p_s)$.
For each $s,k$ we have $\frac{\partial f}{\partial \alpha_{s,k}} = \lambda_s$ when $\alpha_{s,k} >0$
and $\frac{\partial f}{\partial \alpha_{s,k}} \leq \lambda_s$ when $\alpha_{s,k} =0$.
With these conditions, Theorem 4.5.1
in \cite{ref:Gallager} is generalized as follows. We define:
\begin{equation}\label{eq:MutuInfonInput_1}
I(\mathbf{X}=\mathbf{x}_{s,k}; \mathbf{Y})=\sum_{\mathbf{y} \in
{\cal Y}} p(\mathbf{y}|\mathbf{x}_{s,k}) \log
\frac{p(\mathbf{y}|\mathbf{x}_{s,k})}{\sum_{s,k} \alpha_{s,k}
p(\mathbf{y}|\mathbf{x}_{s,k})}
\end{equation}
The necessary
and sufficient conditions for an input probability vector
${\boldsymbol \alpha} \in
{\cal P_{X,S}}$ to maximize this mutual information are state as follows. For some set of numbers $\{C_s\}$, where
$s \in {\cal S}$: If $ \alpha_{s,k} >0$ then $I(\mathbf{X}=\mathbf{x}_{s,k}; \mathbf{Y}) = C_s$; otherwise, if $\alpha_{s,k} =0$ then $I(\mathbf{X}=\mathbf{x}_{s,k}; \mathbf{Y}) \leq  C_s$. Let ${\cal Y}_A$ be the set of all $\mathbf{y}$ whose elements are permutations of a certain $\mathbf{y}_A$. The
$K_s \times |{\cal Y}_A|$ sub--matrix that contains
$p(\mathbf{y}|\mathbf{x}_{s,k})$ which correspond to the inputs
from the state $S=s$ and the outputs from the subset ${\cal Y}_A$
exhibits a symmetry: each row of this sub--matrix is a permutation
of each other row. Using the definition of symmetric channel from \cite{ref:Gallager} and setting all the inputs
$\mathbf{x} \in {\cal X}_s$ equiprobable with
$\alpha_{s,k}=\frac{p_s}{K_s}$. Then $p(\mathbf{y})=\sum_s \frac{p_s}{K_s} \sum_{k} p(\mathbf{y}|\mathbf{x}_{s,k})$, one can check that $I(\mathbf{X}=\mathbf{x}_{s,k}; \mathbf{Y})=C_s$ is constant for
all inputs that belong to the same state $s$.
\end{proof}

\emph{Proof}: The members on the left-handed side of
\eqref{eq:lemmaconcavityQuv_1} can be written as:
{\small \begin{eqnarray}\label{eq:lemmaconcavityQuv_2}
    H\big( (Q_1+Q_2) \mathbf{q}_0 + (Q_3+Q_4) \mathbf{q}_1 \big)
    =
    H\big( \lambda \mathbf{v}_1 + (1\textrm{-}\lambda) \mathbf{v}_2 \big)
    \qquad
    H\big( (Q_1+Q_3) \mathbf{q}_0 + (Q_2+Q_4) \mathbf{q}_1 \big)
    =
    H\big( (1\textrm{-}\lambda) \mathbf{v}_1 + \lambda \mathbf{v}_2 \big) \nonumber
\end{eqnarray}}
where $\mathbf{v}_1 = Q_1 \mathbf{q}_0 + (Q_2+Q_3+Q_4) \mathbf{q}_1$,
$\mathbf{v}_2= (Q_1+Q_2+Q_3) \mathbf{q}_0 + Q_4 \mathbf{q}_1$, and
$\lambda=\frac{Q_3}{Q_2+Q_3}$. Since $H(\cdot)$ is concave, we finalize the proof by writing:
{\small \begin{eqnarray}\label{eq:lemmaconcavityQuv_3}
    H\big( \lambda \mathbf{v}_1 + (1\textrm{-}\lambda) \mathbf{v}_2 \big)+
    H\big( (1\textrm{-}\lambda) \mathbf{v}_1 + \lambda \mathbf{v}_2 \big)
    \geq
    \lambda H(\mathbf{v}_1)+(1\textrm{-}\lambda)H(\mathbf{v}_2)+
    (1\textrm{-}\lambda)H(\mathbf{v}_1)+\lambda H(\mathbf{v}_2 ) =
    H(\mathbf{v}_1)+H(\mathbf{v}_2)  \nonumber
\end{eqnarray}}

\vspace{-16pt}

\subsection{Proof of Theorem~\ref{TheoremMinimalMultisymbol}}
\label{sec:ProofTheoremMinimalMultisymbol}

\begin{IEEEproof}
Let the basic multisymbol associated with $T=t$ be
represented by a matrix:
\begin{equation}\label{eq:basic_multisymbol_matrix}
M = \bordermatrix{~ & \mathbf{z}_1 & \mathbf{z}_2 & \mathbf{z}_3 &
                        \cdots & \mathbf{z}_{F-1} & \mathbf{z}_F\cr
            \mathbf{x}_0 & 0 & 0 & 0 & \cdots & 0 & 0 \cr
            \mathbf{x}_1 & 0 & 0 & 0 & \cdots & 0 & 1 \cr
                 \vdots &       \vdots  & \vdots  & \ddots & \vdots & \vdots
                 \cr
            \mathbf{x}_{F-1} & 1 & 1 & 1 & \cdots & 1 & 0 \cr
            \mathbf{x}_F & 1 & 1 & 1 & \cdots & 1 & 1}
\end{equation}
It can be easily checked that for any pair $\mathbf{z}_{f_1},
\mathbf{z}_{f_2}$ either the set ${\cal G}_{10}(z_1,z_2)$ or the
set ${\cal G}_{01}(z_1,z_2)$ is empty. According to
Lemma~\ref{lemma:Empty_Sets}, that permutation (swapping) of the
values within one or more $\mathbf{x}_s$ cannot further decrease
the entropy contribution of the positions that are swapped. Hence,
the basic multisymbol \eqref{eq:basic_multisymbol_matrix} results
in the minimal possible value of $H(\mathbf{Y}|{\cal M}_t)$.
The same observation can be made whenever ${\cal M}_t$ is a minimal multisymbol,
which proves the theorem.
\end{IEEEproof}

%

\subsection{Proof of Theorem~\ref{Theorem_Possible_choice}}
\label{sec:ProofTheorem_Possible_choice}

Since $L$ divides each $\binom{F}{s}$, the number of multisymbols
that contain $\mathbf{x}_s \in {\cal X}_s$ is an integer
$m_s=\frac{L}{\binom{F}{s}}$. The number of outgoing edges from
$\mathbf{x}_s$ is $(F-s)$, while the number of incoming edges to
$\mathbf{x}_s$ is $s$. The sum of incoming weights and the sum of
outgoing weights for $\mathbf{x}_s$ is equal to $m_s$. Note that
the average outgoing weight for $\mathbf{x}_s$ is
$\frac{m_s}{F-s}$, while the average incoming weight for any
$\mathbf{x}_{s+1} \in {\cal X}_{s+1}$ is $\frac{m_{s+1}}{s+1}$.
However, the following holds $\frac{m_s}{F-s}=\frac{L}{\binom{F}{s}
(F-s)}=\frac{L}{\binom{F}{s+1}(s+1)}=\frac{m_{s+1}}{s+1}$
i. ~e. the average outgoing weight from ${\cal X}_s$ is equal to
the average incoming weight at ${\cal X}_{s+1}$, which is a
necessary condition for the multisymbols that achieve the secondary capacity. We now prove that for each outgoing weight from
$\mathcal {X}_s$ there is a matched incoming weight at
$\mathcal{X}_{s+1}$.We choose the weight of each edge to be either
$w_1=\lfloor\frac{m_{s}}{F-s}\rfloor$ or
$w_2=\lceil\frac{m_{s}}{F-s}\rceil$. Then $b$ weights have to be chosen to be equal to
$w_2=\lceil\frac{m_s}{F-s}\rceil$, where $b$ is given by
\begin{equation}
m_{s}=a(F-s)+b, \: a\in \{\mathbb{N}\cup 0\},\:0\leq b\leq F-s-1.
\label{eq:remainder1}
\end{equation}
There are $s+1$ incoming edges at
$\mathbf{x}_{s+1}$. The weight of each incoming edge is also either
$w_1$ or $w_2$, since $\frac{m_{s}}{F-s}=\frac{m_{s+1}}{s+1}$. In
order to satisfy the condition that the total incoming weight of
$\mathbf{x}_{s+1}$ is $m_{s+1}$, $d$ weights should be chosen
to be equal to $w_2$, where $d$ is given by
\begin{equation}
m_{s+1}=c(s+1)+d, \:c\in\{\mathbb{N}\cup 0\},\:0\leq d \leq s.
\label{eq:remainder2}
\end{equation}
If (\ref{eq:remainder1}) and (\ref{eq:remainder2}) are satisfied, then
$b\binom{F}{s}=d\binom{F}{s+1}$ needs to be fulfilled, which
follows from $\binom{F}{s+1}=\binom{F}{s}\frac{F-s}{s+1}$ and the equality of average incoming/outgoing weights.
For each outgoing weight from ${\cal X}_s$ there is a
matched incoming weight at ${\cal X}_{s+1}$. Since
$L \leq F!$, it will be always possible to select $L$ different
paths.

\bibliographystyle{IEEEtran}
\bibliography{references}

\end{document}